\documentclass[nohyperref]{article}

\usepackage{microtype}
\usepackage{graphicx}
\newcommand{\nblink}[1]{\href{https://github.com/DifferentiableUniverseInitiative/jaxpm-paper/blob/v_icml/notebooks/#1.ipynb}{\faFileCodeO}}
\newcommand{\github}{\href{https://github.com/DifferentiableUniverseInitiative/jaxpm-paper/tree/v_icml}{\faGithub}}
\usepackage{fontawesome}
\usepackage{subfigure}
\usepackage{booktabs} 
\usepackage{hyperref}

\usepackage[accepted]{icml2022}
\usepackage{amsmath}
\usepackage{amssymb}
\usepackage{mathtools}
\usepackage{amsthm}
\usepackage{soul}

\usepackage[capitalize,noabbrev]{cleveref}

\theoremstyle{plain}

\theoremstyle{definition}

\theoremstyle{remark}

\usepackage[textsize=tiny]{todonotes}

\icmltitlerunning{Hybrid Physical-Neural ODEs for Fast N-body Simulations}

\begin{document}

\twocolumn[
\icmltitle{Hybrid Physical-Neural ODEs for Fast N-body Simulations}
\icmlsetsymbol{equal}{*}

\begin{icmlauthorlist}
\icmlauthor{Denise Lanzieri}{equal,yyy}
\icmlauthor{François Lanusse}{equal,xxx}
\icmlauthor{Jean-Luc Starck}{xxx}

\end{icmlauthorlist}

\icmlaffiliation{yyy}{Université Paris Cité, Université Paris-Saclay, CEA, CNRS, AIM, F-91191, Gif-sur-Yvette, France}
\icmlaffiliation{xxx}{Université Paris-Saclay, Université Paris Cité, CEA, CNRS, AIM, 91191, Gif-sur-Yvette, France; }

\icmlcorrespondingauthor{Denise Lanzieri}{denise.lanzieri@cea.fr}

\icmlkeywords{Machine Learning, ICML}

\vskip 0.3in
]

\printAffiliationsAndNotice{\icmlEqualContribution}

\begin{abstract}
We present a new scheme to compensate for the small-scales approximations resulting from Particle-Mesh (PM) schemes for cosmological N-body simulations. This kind of simulations are fast and low computational cost realizations of the large scale structures, but lack resolution on small scales.
To improve their accuracy, we introduce an additional effective force within the differential equations of the simulation, parameterized by a Fourier-space Neural Network acting on the PM-estimated gravitational potential. 
We compare the results for the matter power spectrum obtained to the ones obtained by the PGD scheme (Potential Gradient Descent scheme).
We notice a similar improvement in term of power spectrum, but we find that our approach outperforms PGD for the cross correlation coefficients, and is more robust to changes in simulation settings (different resolutions, different cosmologies). 
We provide all codes associated with this paper at this link \github.
\end{abstract}

\section{Introduction}

N-body simulations have become an essential tool for the cosmological  analysis of current and upcoming sky surveys. Because solving the full N-body problem can be very expensive in terms of computational resources and time, Quasi N-body numerical schemes have been proposed to speed up the computational time and create low cost realizations of the large scale structure, e.g. FastPM \cite{feng2016fastpm} and COLA \cite{tassev2013solving}. 
In this work, we adopt a Particle-Mesh (PM) scheme, which approximates the computation of the gravitational forces between the dark matter particles, by computing the gravitational potential on a 3D mesh. The drawback of this PM scheme is that it approximates the short range interactions between particles and thus is not able to resolve structures with scales smaller than the mesh resolution. As a result, these simulations typically lack power on small scales and halo profiles are less sharp than their full N-body counterparts.

In this work, we propose to improve the accuracy of PM simulations by augmenting the physical differential equations of a PM N-body with a minimally-parametric neural network component modeling a \textit{residual effective force} compensating for the PM approximations. In order to impose rotational and translational invariance, this neural correction is implemented in the form of a learned isotropic Fourier filter.

To train the neural network through the simulator, we follow the scheme proposed by \citet{chen2018neural}, i.e. we treat the Ordinary Differential Equations (ODE) solver running the simulation as a black box and compute gradients using the adjoint sensitivity method \cite{pontriagin1964mathematical}.
We train and compare the model to the CAMELS simulations \cite{villaescusa2021camels} and test it against several changes of the simulation setting (resolution, volume, cosmological parameters). We also compare our method to PGD approach (Potential Gradient Descent
scheme) \cite{dai2018gradient}.

\section{Related work}
Several machine learning techniques are available to emulate high cost N-body simulations. In particular, \citet{he2019learning} proposed a deep neural network to learn the nonlinear mapping from first order perturbation theory linear displacements to the displacement field of FastPM simulations. More recently, \citet{li2021ai} have demonstrated that a generative model can be trained to super-resolve the particle displacement field, allowing to enhance the resolution of a low cost approximate N-body.
These approaches use particle displacements as inputs and outputs of their modeling, but rely on large Deep Convolutional Networks to learn an effective mapping generating the desired outputs.

\paragraph{FastPM}
FastPM \citep{feng2016fastpm} is a quasi particle-mesh (PM) N-body solver, using a specifically designed leapfrog integration scheme modified such that it agrees with the Zeldovich solution on large scales. In this way, at large scale (k$ \to 0$), the correct linear theory growth can be produced in a limited number of steps. 

\paragraph{Potential Gradient Descent (PGD)}
\citet{dai2018gradient} introduces a gradient based method to correct the PM approximation in FastPM and improve the modeling of the matter distribution within halos.
PGD models the effect of short range interactions as an additional particle displacement term, moving the particles towards a minimum of the gravitational potential after band-pass filtering: 
\begin{equation}\label{displacement}
    \textbf{S}=
    \frac{4\pi G \bar{\rho}\alpha_{PM}}{H_0^2}
    \nabla
   \hat{\textbf{O}}_l(k)
     \hat{\textbf{O}}_s(k)
    \nabla^{-2}\delta,
\end{equation}
where $\hat{\textbf{O}}_l(k)= \exp{(\frac{k^2}{k^4_l})}$ and $\hat{\textbf{O}}_s(k)= \exp{(-\frac{k^4}{k^4_s}})$ are the low and high pass filter introduced to remove the long range force and to reduce the numerical effects induced by the mesh resolution. 
PGD introduces 3 nuisance parameters fitted on training simulations: $\alpha_{PM}$, defining the amplitude of the filter and the long and the short range scale parameters $k_l$ and $k_s$.

\section{Hybrid Physical-Neural ODE}

Cosmological simulations implement integration of gravitational evolution starting from the Gaussian initial conditions of the Universe to the observed large scale structures. \\
 In this work, we choose to compute the time integration starting from a system of Ordinary Differential Equations (ODE) and treat the ODE solver as a black box:
\begin{equation}
    \left\{ \begin{array}{ll}
        \frac{d \mathbf{x}}{d a} & = \frac{1}{a^3 E(a)} \mathbf{v} \\
        \frac{d \mathbf{v}}{d a} & =  \frac{1}{a^2 E(a)} F_\theta(\mathbf{x}, a), \\
    \end{array} \right.
\end{equation}
where $\mathbf{x}$ and $\mathbf{v}$ define the position and the velocity of the particle and \textit{a} is the cosmological scale factor.
We have also introduced the parametric function $F_\theta$, hybrid between a physical model and a neural network:
\begin{equation}\label{hybrid_model}
    F_\theta(\mathbf{x}, a) = \frac{3 \Omega_m}{2}  \nabla \left[ \phi_{PM} (\mathbf{x}) \ast \mathcal{F}^{-1}(1 + f_\theta(a,|\mathbf{k}|)) \right],
\end{equation}

where $\mathcal{F}^{-1}$ is the inverse Fourier transform, $\phi_{PM}$ is the gravitational potential estimated by using the cloud-in-cell (CiC) interpolation scheme \cite{hockney2021computer} and $f_\theta(a,|\mathbf{k}|)$ is the learned neural filter, aimed to model the residual effective force compensating for the PM approximations. We choose to define $f_\theta(a,|\mathbf{k}|)$ as a B-spline function, whose coefficients (knot points) are the output of a neural network of parameters $\theta$.
More specifically, we choose B-spline of order 3, defined over 16 knot points. 
We point out that, since we are modeling true physical processes, there are several symmetries we wish to preserve. Therefore, $f_\theta$ is implemented as a Fourier-based isotropic filter in order to preserve the translational and rotational symmetries.

To train our hybrid physical-neural ODE and back-propagate through the ODE solver, we compute the gradient adopting the adjoint sensitivity method \cite{chen2018neural, pontryaginmathematical}, consisting in solving a second ODE backwards in time, and treat the ODE solver as a black box.  We adopt the following loss function penalizing both the positions of the particles and the overall matter power spectrum at different snapshot times ${s}$, compared to a reference full N-body simulation:

\begin{equation}\label{loss2}
    \mathcal{L} =  \sum_{s} || \mathbf{x}^{Nbody}_s - \mathbf{x}_s ||_2^2  + \lambda || \frac{P_s(k)}{P_s^{Nbody}(k)} -1 ||_2^2 \; ,
\end{equation}
where $P_s(k)$ is the power spectrum of the simulation for snapshot $s$, and $\lambda$ is left as hyper-parameter balancing the contributions of the two terms.
In contrast to \citet{he2019learning} who train the model minimizing the mean square error on particle displacements, we empirically found that penalizing the overall matter power spectrum leads to a better improvement for the small scale power spectrum, compared to penalizing the particle positions only. 
One possible reason for this result, may be the reduced number of degrees of freedom in our model compared to the one used by \citet{he2019learning}, i.e. in our training process, the neural network cannot approximate arbitrary non-linear functions, it is restricted by physical constraints and our simple parameterisation. 
However, we know that different displacement fields could produce identical density fields and then, identical power spectrum. If we penalised the power spectrum term only, in principle we may increase the overall power spectrum on small scales, without however recovering the right positions of the particles. For this reason, the final loss function we use includes both terms of \autoref{loss2}. 
We empirically find $\lambda=0.1$ provides good results. The model is trained by gradient descent with the Adam optimizer \cite{kingma2014adam} and learning rates of 0.01.
The code is implemented in the Python package Jax \cite{jax2018github} and we use the Jax-based Dormand-Prince ODE solver. 

\section{Simulation data}
In this work, we rely on the CAMELS dataset \cite{villaescusa2021camels}, in particular we use the suite of dark-matter only N-body simulations implementing the N-body tree-particle-mesh approach of IllustrisTNG \cite{nelson2019illustristng}. 
We have 34 snapshots, generated following the evolution of $256^3$ dark matter particles in a periodic box of comoving volume equal to $25^3$ ($h^{-1}$ Mpc)$^3$, with initial conditions generated at z=127 using second order Lagrangian perturbation theory (2LPT). 
The following cosmological parameters are kept fixed in all simulations:  $h=0.6711$, $n_s =0.9624$, $M_{\nu} = 0.0$ eV, $w=-1$, $\Omega_k = 0.$, while the values of $\Omega_m$ and $\sigma_8$ are varied across simulations.  We further downsample the CAMELS particle data, down to $64^3$ particles, to make the problem more manageable on a single GPU.

\section{Results}\label{results}
\begin{figure}
    \centering
    \includegraphics[width=\columnwidth]{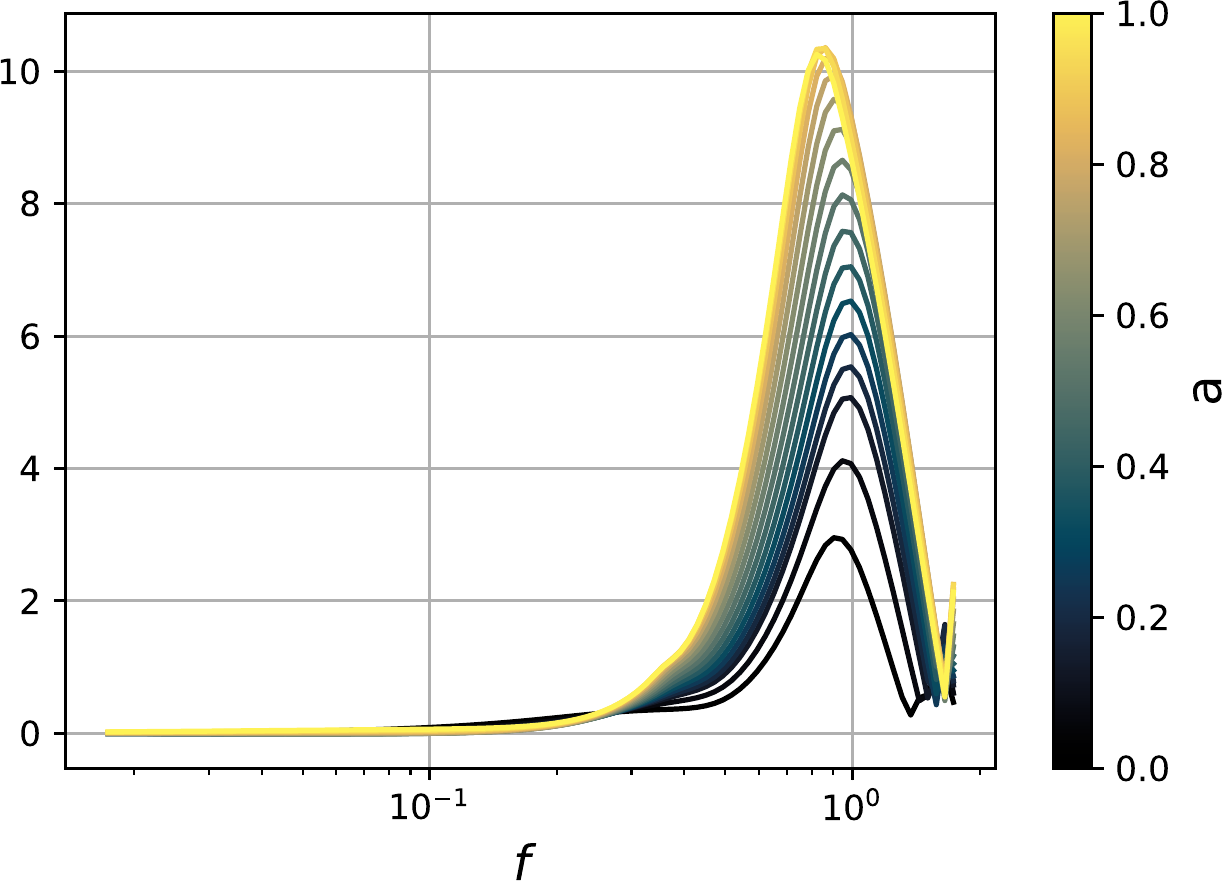}
    \caption{Learned neural filter $f_\theta(a, |\mathbf{k}|)$ as a function of normalized mesh frequency ($f=1$ corresponds to the Nyquist frequency of the mesh), for different scale factor \textit{a}.}
    \label{fig:filter}
\end{figure}
\begin{figure*}
    \centering
    \includegraphics[width=\textwidth]{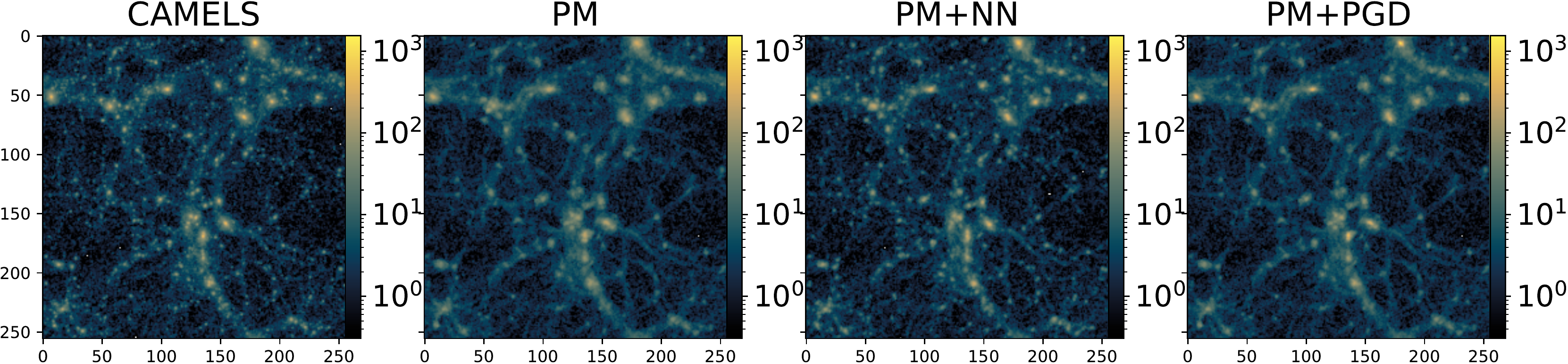}
    \caption{From left to right: Projections of final density field for CAMELS, PM only, PM corrected by the NN scheme and PM corrected by PGD. Although the PGD model improves the sharpness of the halos, it fails to fully recreate the dark matter structures,  producing a smoother density profile compared to the NN model, which appears closer to CAMELS. \nblink{NNvsPGD_CV_0} }
    \label{fig:cluster_diff_schemes}
\end{figure*}

In this section, we train and evaluate our hybrid neural ODE on CAMELS data, and compare it to results obtained with the PGD approach of \citet{dai2018gradient}. We train the neural ODE using the loss function presented in \autoref{loss2}, and the PGD parameters are fitted only for the final snapshot ($a=1$) by penalizing the weighted power spectrum ratio between the corrected and reference simulations (i.e. \autoref{lossPGD}). 

In both cases, we use a \textit{single} CAMELS N-body simulation at the fiducial cosmology of $\Omega_m=0.3$ and $\sigma _8=0.8$ to fit the parameters of the models. 

\autoref{fig:cluster_diff_schemes} shows the matter overdensity fields obtained with the 2 different correction schemes compared to the CAMELS and the standalone PM (i.e. pure PM simulation). As we will quantify below, we can already see that the pure PM simulation (second panel from the left) is smoother than the reference CAMELS simulation, and that the neural correction (third panel from the left) recovers most of the missing information. By comparison, the PGD correction does sharpen somewhat some structures in the field, in order to increase the overall power spectrum on small scales, but is less successful at restoring the actual shape of the dark matter structures.

\begin{figure}
    \centering
    \includegraphics[width=\columnwidth]{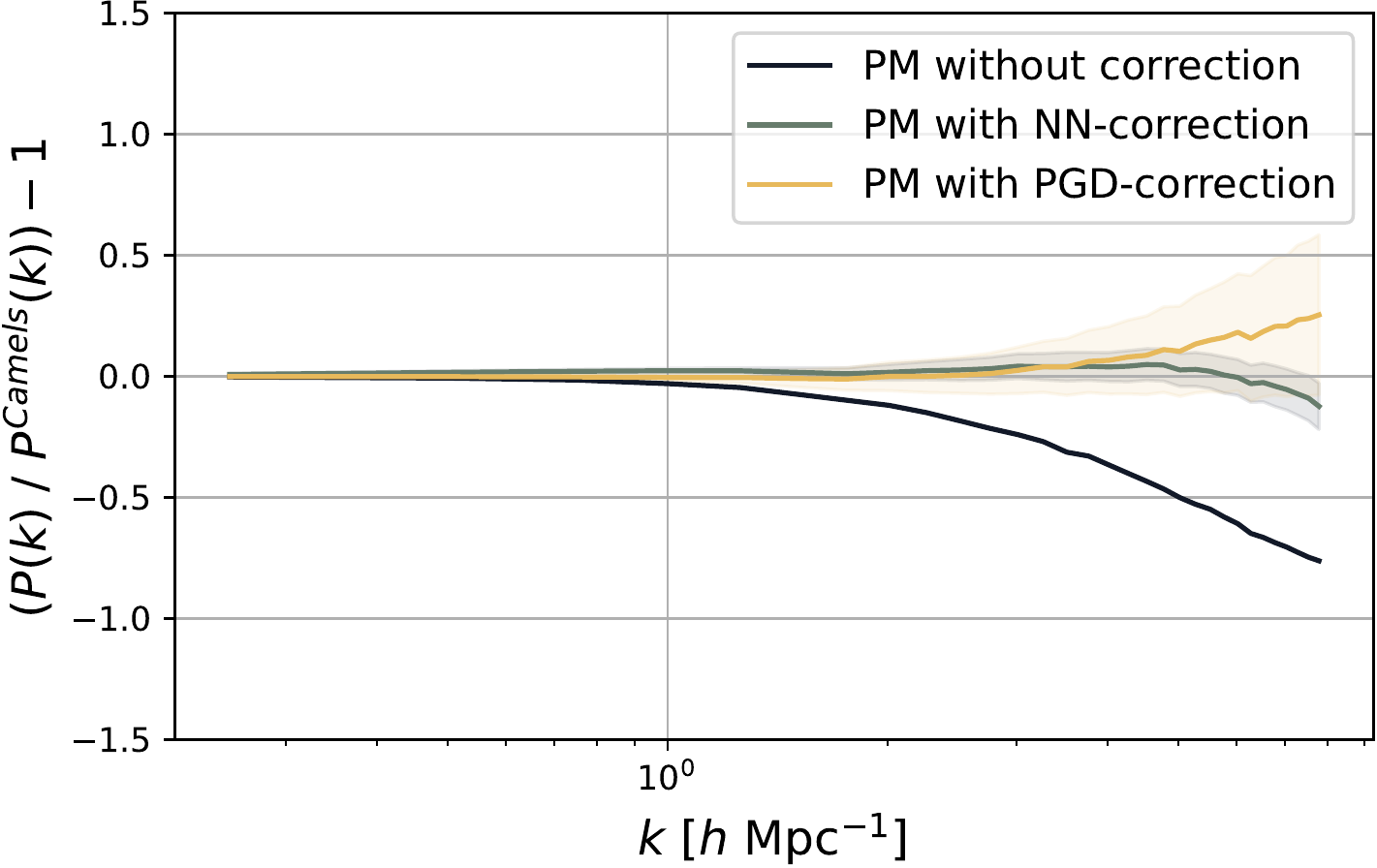}
    \includegraphics[width=\columnwidth]{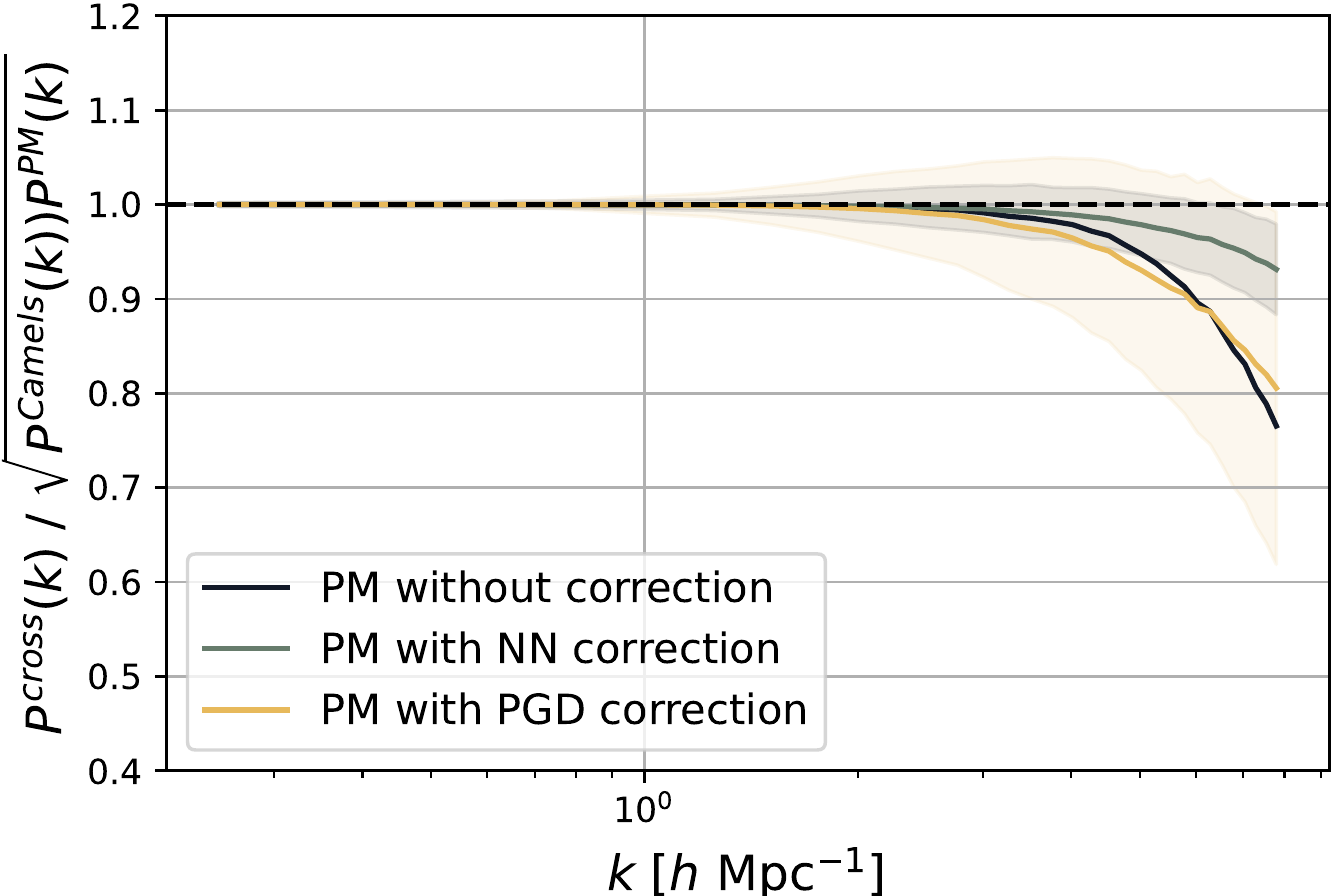}
    \caption{ Top panel:
    Fractional matter power spectrum of PM simulations before and after using the correction models and CAMELS simulations. Bottom panel: cross correlation coefficients with the reference simulation before and after the two different correction schemes.
    The shaded regions represent the standard deviation from the same realisation corrected with 10 different parameters fitted from different initial conditions. 
    \nblink{NNvsPGD_CV_0}}
    \label{fig:res_same_params}
\end{figure}

\paragraph{Learned Neural Filter} In \autoref{fig:filter} we show the trained Fourier-space filter $f_\theta(a, |\mathbf{k}|)$ as a function of normalized mesh frequency. As expected, the correction only affects small scales, having no effect on the large-scale modes of the simulations for which the PM solver is accurate. An interesting feature to note is that the correction is scale-factor dependent. This indicates that the model is not simply applying a sharpening filter akin to a CiC compensation filter, but that the correction most likely adapts to the density field properties at different cosmological times. One drawback of this observation though is that if the neural correction is dependent on the particular dynamics of the dark matter density field, it may be cosmology-dependent. 
\paragraph{Power spectrum comparison at fiducial cosmology} 
We show on \autoref{fig:res_same_params} a comparison of the ratio of the power spectrum and the cross-correlation of CAMELS and PM simulations to illustrate the difference between the two correction schemes. As can be seen, while most of the missing power is recovered on small scales in both cases, the improvement of cross correlation coefficients after using our scheme is drastically different.

\paragraph{Robustness to changes in resolution and cosmological parameters} To evaluate the robustness of our correction scheme, we compute the following tests:
\begin{enumerate}
    \item  
    We investigate the effects due to \textbf{varying the cosmological parameters}. We run the PM simulations and compare them against a set of CAMELS, both generated at $\Omega_m= 0.10$ (instead of $\Omega_m= 0.3$ used for the training) and the same initial conditions. The top panel of \autoref{fig:diff_sett} presents the result of this test. Between PGD and NN correction schemes, the NN correction shows a slight deviation compared to results obtained at the fiducial cosmology on \autoref{fig:res_same_params}. PGD, on the contrary, appears to be extremely sensitive to $\Omega_m$, no longer applying the desired correction.
 \begin{figure}
    \centering
    \includegraphics[width=\columnwidth]{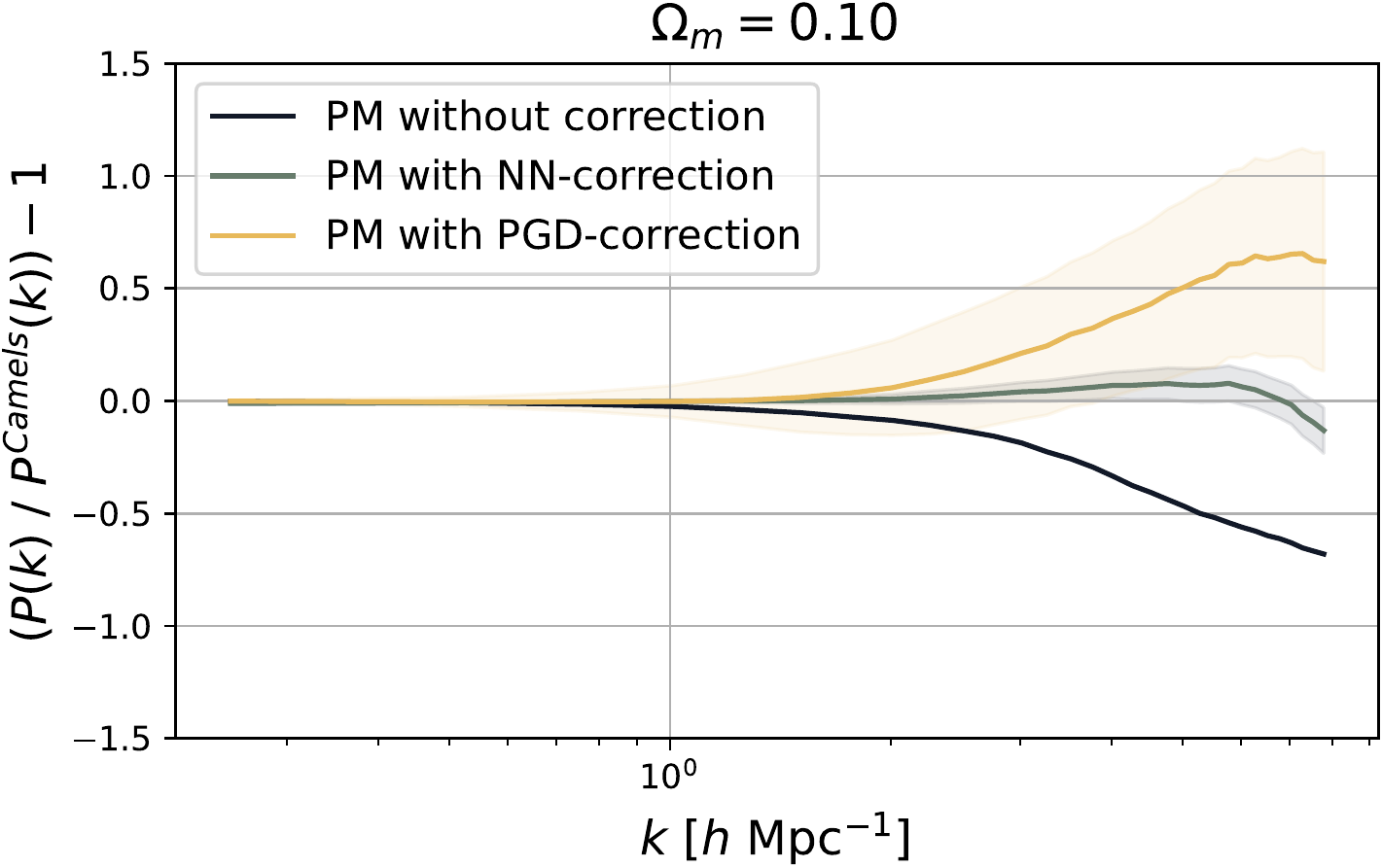}
    \includegraphics[width=\columnwidth]{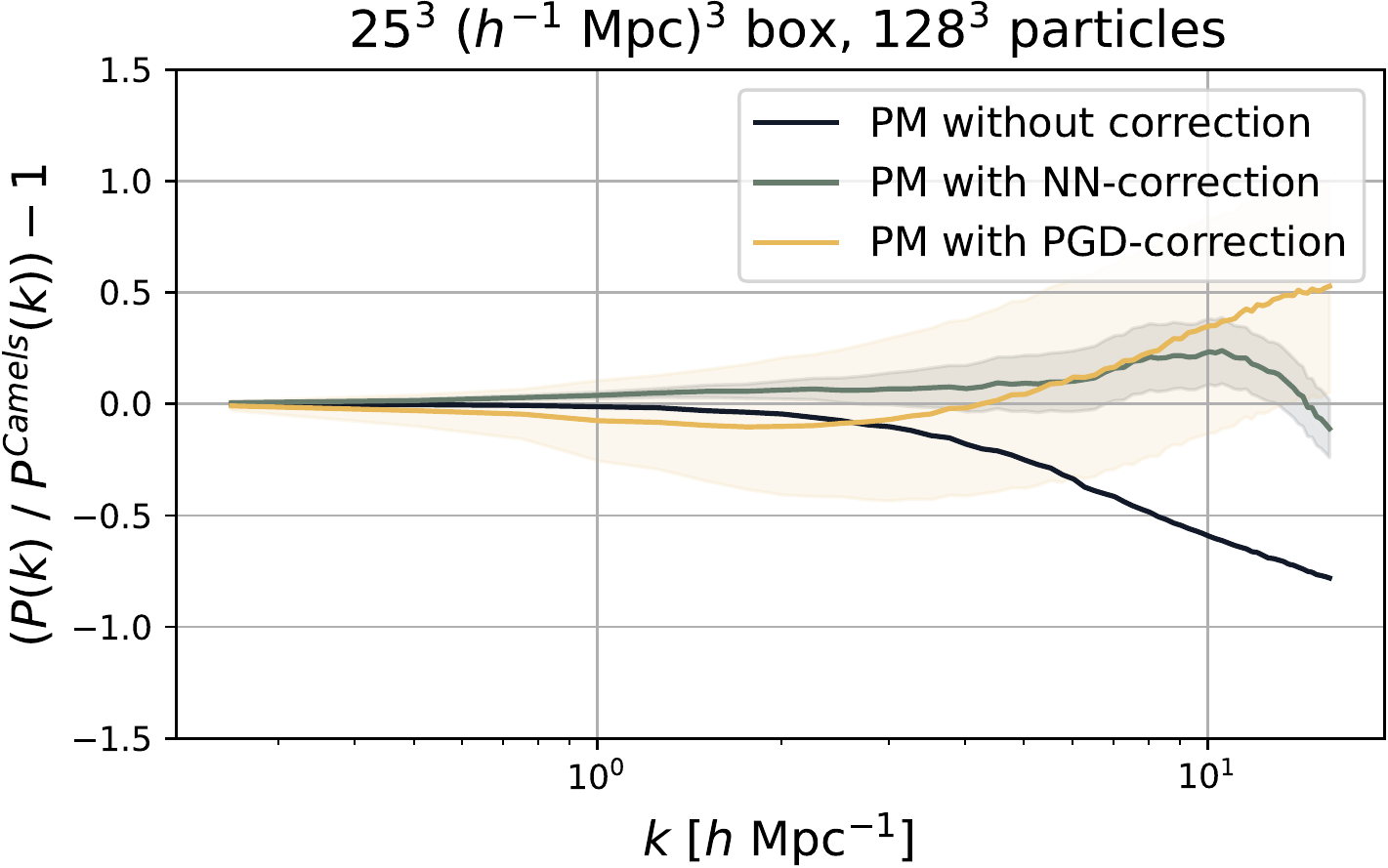}
    \includegraphics[width=\columnwidth]{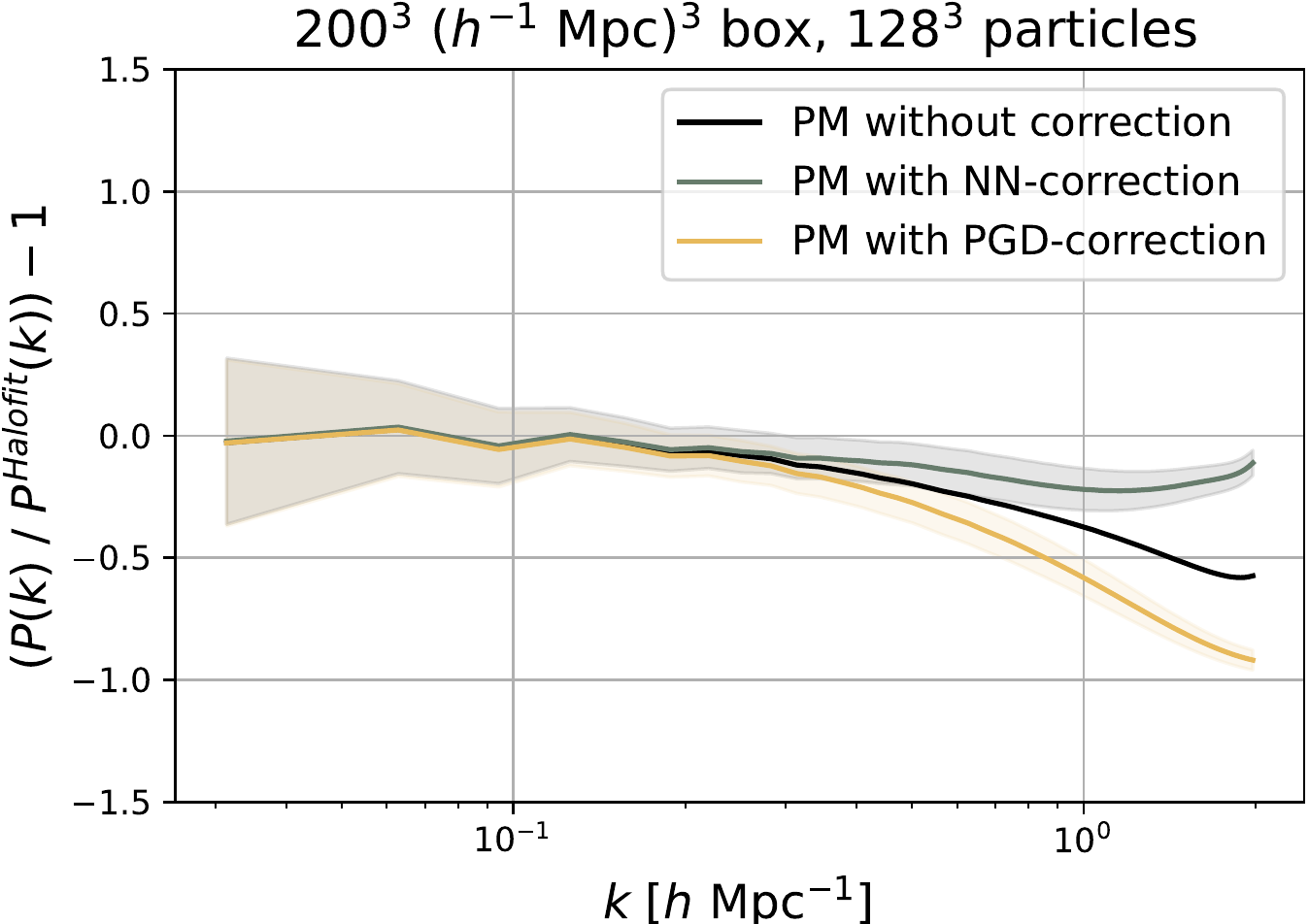}
    \caption{ Benchmarks on matter power spectrum, varying the setting of the simulations. Top panel: effects due to varying the cosmological parameters $\Omega_m$ employed in the simulation \nblink{NNvsPGD_wrong_cosmo_omega_M}.
    Center panel: effects due to increasing the number of particles (to $128^3$) employed in the simulation \nblink{NNvsPGD_wrong_resolution}. 
    Bottom panel: effects due to increasing the number of particles ($128^3$) and box size (205 $h^{-3}$Mpc$^3$) employed in the simulation. Note that in this last comparison the theoretical HaloFit matter power spectrum is considered as reference \nblink{NNvsPGD_wrong_boxsize_resolution}.
    In the top and central panel the shaded regions represent the standard deviation from the same realisation corrected with 10 different parameters fitted from different initial conditions. In the bottom panel the shaded regions represent the standard deviation from 1000 realisations with different initial conditions corrected with the same parameters.  }
    \label{fig:diff_sett}
\end{figure}
 \item 
 We investigate the effects of \textbf{increasing the resolution of the simulations}. We run the PM simulation with $128^3$ particles for the same 25 $h^{-1} Mpc$ box size. The center panel of \autoref{fig:diff_sett} shows that the NN correction out-performs PGD up to $k \sim 7$, after which the method overemphasizes the small scale power. 

 \item 
 Finally, we test the two schemes at \textbf{lower resolution}, increasing the volume of the simulation to 200 ($h^{-1}$Mpc)$^3$ and the number of particles to $128^3$. Note that the CAMELS simulation are produced in a box of 25 ($h^{-1}$ Mpc)$^3$, therefore in this test we take as reference the halofit power spectrum $P(k)/P(k)^{halofit}$.
  Based on the bottom panel of \autoref{fig:diff_sett}, we can see that, even in this extreme case, the performance of the NN correction remains very good, while PGD no longer applies the desired correction. 

 \end{enumerate}

\section{Conclusion}

We have presented in this work an alternative to the PGD correction scheme for a quasi N-body PM solver, based on Neural Network implemented as a Fourier-space filter. We illustrated the merits of this approach by comparing the results of a corrected PM simulations against the high resolution CAMELS simulations.  
We benchmarked our model against the PGD scheme, showing that, at fiducial cosmology, the two methods are able to give similar improvement to the small scale power spectrum, but significantly different improvement in the correlation coefficients. We have also presented the results to changes in resolution and cosmological parameters, proving that our method gives better results for both the power spectrum and cross correlation coefficients, it is therefore less sensitive to the setting of the simulations used for the training. We note however, that our scheme slightly overemphasizes the small scale power for $k \sim 7$ and is not able to strongly improve the results for the cross correlation coefficients when tested to simulations of higher resolutions. 

\newpage
\bibliography{example_paper}

\begin{thebibliography}{13}
\providecommand{\natexlab}[1]{#1}
\providecommand{\url}[1]{\texttt{#1}}
\expandafter\ifx\csname urlstyle\endcsname\relax
  \providecommand{\doi}[1]{doi: #1}\else
  \providecommand{\doi}{doi: \begingroup \urlstyle{rm}\Url}\fi

\bibitem[Bradbury et~al.(2018)Bradbury, Frostig, Hawkins, Johnson, Leary,
  Maclaurin, Necula, Paszke, Vander{P}las, Wanderman-{M}ilne, and
  Zhang]{jax2018github}
Bradbury, J., Frostig, R., Hawkins, P., Johnson, M.~J., Leary, C., Maclaurin,
  D., Necula, G., Paszke, A., Vander{P}las, J., Wanderman-{M}ilne, S., and
  Zhang, Q.
\newblock {JAX}: composable transformations of {P}ython+{N}um{P}y programs,
  2018.
\newblock URL \url{http://github.com/google/jax}.

\bibitem[Chen et~al.(2018)Chen, Rubanova, Bettencourt, and
  Duvenaud]{chen2018neural}
Chen, R.~T., Rubanova, Y., Bettencourt, J., and Duvenaud, D.~K.
\newblock Neural ordinary differential equations.
\newblock \emph{Advances in neural information processing systems}, 31, 2018.

\bibitem[Dai et~al.(2018)Dai, Feng, and Seljak]{dai2018gradient}
Dai, B., Feng, Y., and Seljak, U.
\newblock A gradient based method for modeling baryons and matter in halos of
  fast simulations.
\newblock \emph{Journal of Cosmology and Astroparticle Physics}, 2018\penalty0
  (11):\penalty0 009, 2018.

\bibitem[Feng et~al.(2016)Feng, Chu, Seljak, and McDonald]{feng2016fastpm}
Feng, Y., Chu, M.-Y., Seljak, U., and McDonald, P.
\newblock Fastpm: a new scheme for fast simulations of dark matter and haloes.
\newblock \emph{Monthly Notices of the Royal Astronomical Society},
  463\penalty0 (3):\penalty0 2273--2286, 2016.

\bibitem[He et~al.(2019)He, Li, Feng, Ho, Ravanbakhsh, Chen, and
  P{\'o}czos]{he2019learning}
He, S., Li, Y., Feng, Y., Ho, S., Ravanbakhsh, S., Chen, W., and P{\'o}czos, B.
\newblock Learning to predict the cosmological structure formation.
\newblock \emph{Proceedings of the National Academy of Sciences}, 116\penalty0
  (28):\penalty0 13825--13832, 2019.

\bibitem[Hockney \& Eastwood(2021)Hockney and Eastwood]{hockney2021computer}
Hockney, R.~W. and Eastwood, J.~W.
\newblock \emph{Computer simulation using particles}.
\newblock crc Press, 2021.

\bibitem[Kingma \& Ba(2014)Kingma and Ba]{kingma2014adam}
Kingma, D.~P. and Ba, J.
\newblock Adam: A method for stochastic optimization.
\newblock \emph{arXiv preprint arXiv:1412.6980}, 2014.

\bibitem[Li et~al.(2021)Li, Ni, Croft, Di~Matteo, Bird, and Feng]{li2021ai}
Li, Y., Ni, Y., Croft, R.~A., Di~Matteo, T., Bird, S., and Feng, Y.
\newblock Ai-assisted superresolution cosmological simulations.
\newblock \emph{Proceedings of the National Academy of Sciences}, 118\penalty0
  (19):\penalty0 e2022038118, 2021.

\bibitem[Nelson et~al.(2019)Nelson, Springel, Pillepich, Rodriguez-Gomez,
  Torrey, Genel, Vogelsberger, Pakmor, Marinacci, Weinberger,
  et~al.]{nelson2019illustristng}
Nelson, D., Springel, V., Pillepich, A., Rodriguez-Gomez, V., Torrey, P.,
  Genel, S., Vogelsberger, M., Pakmor, R., Marinacci, F., Weinberger, R.,
  et~al.
\newblock The illustristng simulations: public data release.
\newblock \emph{Computational Astrophysics and Cosmology}, 6\penalty0
  (1):\penalty0 1--29, 2019.

\bibitem[Pontriagin(1964)]{pontriagin1964mathematical}
Pontriagin, L.
\newblock \emph{The Mathematical Theory of Optimal Processes}.
\newblock International series of monographs in pure and applied mathematics.
  Pergamon Press; [distributed in the Western Hemisphere by Macmillan, New
  York], 1964.
\newblock ISBN 9780080101767.
\newblock URL \url{https://books.google.fr/books?id=aakrAAAAYAAJ}.

\bibitem[Pontryagin et~al.(1962)Pontryagin, Boltyanski, Gamkrelidze, and
  Mishchenko]{pontryaginmathematical}
Pontryagin, L., Boltyanski, V., Gamkrelidze, R., and Mishchenko, E.
\newblock The mathematical theory of optimal processes. 1962. interscience.
\newblock \emph{New York}, 1962.

\bibitem[Tassev et~al.(2013)Tassev, Zaldarriaga, and
  Eisenstein]{tassev2013solving}
Tassev, S., Zaldarriaga, M., and Eisenstein, D.~J.
\newblock Solving large scale structure in ten easy steps with cola.
\newblock \emph{Journal of Cosmology and Astroparticle Physics}, 2013\penalty0
  (06):\penalty0 036, 2013.

\bibitem[Villaescusa-Navarro et~al.(2021)Villaescusa-Navarro,
  Angl{\'e}s-Alc{\'a}zar, Genel, Spergel, Somerville, Dave, Pillepich,
  Hernquist, Nelson, Torrey, et~al.]{villaescusa2021camels}
Villaescusa-Navarro, F., Angl{\'e}s-Alc{\'a}zar, D., Genel, S., Spergel, D.~N.,
  Somerville, R.~S., Dave, R., Pillepich, A., Hernquist, L., Nelson, D.,
  Torrey, P., et~al.
\newblock The camels project: Cosmology and astrophysics with machine-learning
  simulations.
\newblock \emph{The Astrophysical Journal}, 915\penalty0 (1):\penalty0 71,
  2021.

\end{thebibliography}
\bibliographystyle{icml2022}


\appendix
\section{Train and validation loss for PGD}
We fit the PGD parameters only for the final snapshot (a=1) by penalizing the following loss function:
\begin{equation}\label{lossPGD}
    \mathcal{L} =  \left\| 
    \sigma(k)\left(  \frac{p(k)}{p^{Nbody}(k)} -1 \right) \right\|_2^2 \; ,
\end{equation}
where:
\begin{equation}
     \sigma(k)=
     \frac
    { 1-\frac{k}{\pi (\frac{\text{mesh shape}}{\text{box volume}})}}
    {\sum_k \left(1-\frac{k}{\pi (\frac{\text{mesh shape}}{\text{box volume}})}\right)}
\end{equation}
is the weight we add to the loss function to downweight the small scales compared to the large scales. We find that this weight choice gives a better improvement in the small scale power spectrum compared to the only power spectrum case. However it doesn’t lead to any improvements or drastic differences to the performance of PGD when applied to simulations with different cosmologies or resolutions.

\section{Cross-Correlation}
\begin{figure}
    \centering
    \includegraphics[width=\columnwidth]{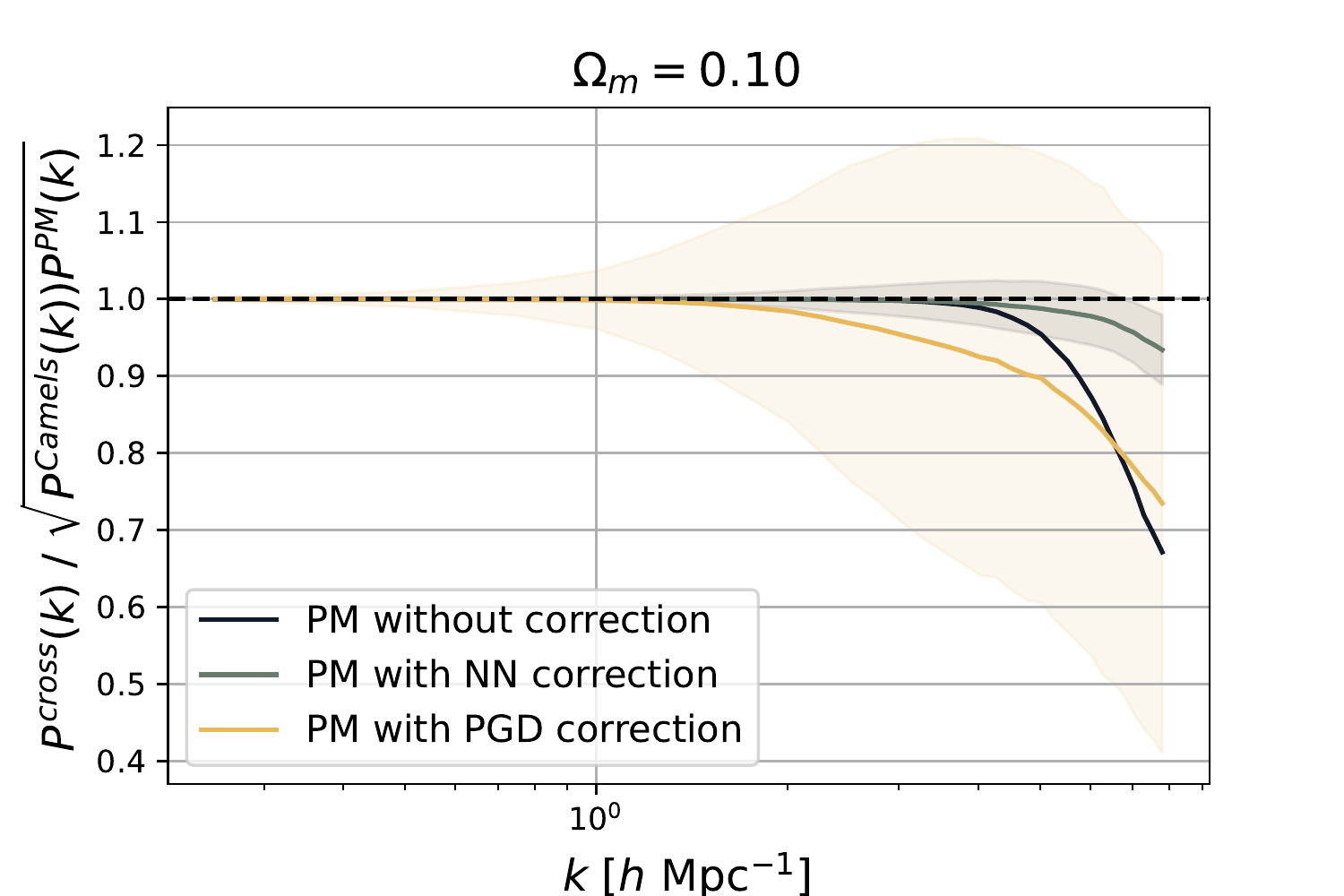}
    \includegraphics[width=\columnwidth]{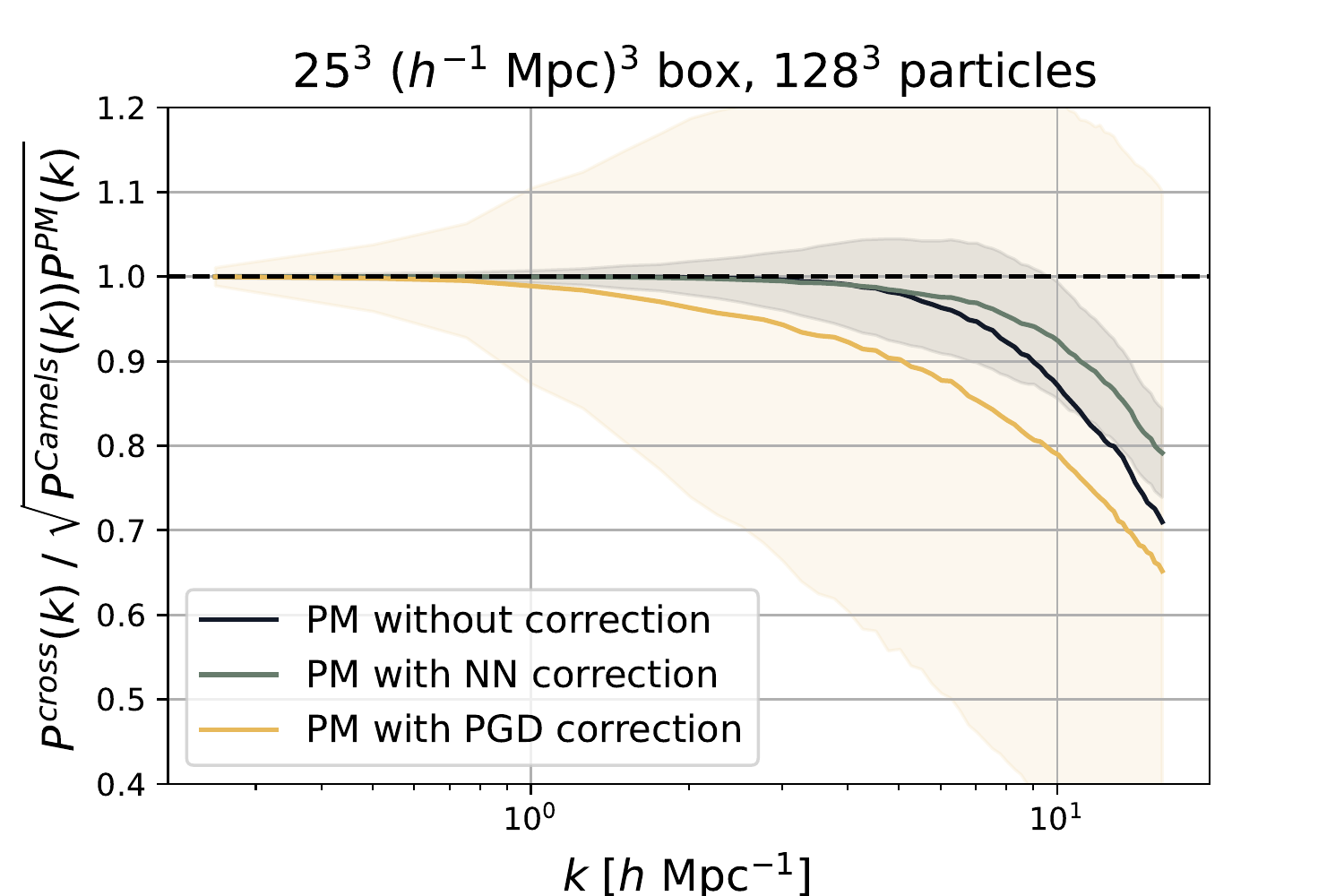}
    \caption{ 
    Cross-correlation of PM and CAMELS simulations, before and after using our models and PGD. Top panel: varying the cosmological parameters $\Omega_m$ employed in the simulation \nblink{NNvsPGD_wrong_cosmo_omega_M}. Bottom panel: varying the number of particles ($128^3$) employed in the simulation \nblink{NNvsPGD_wrong_resolution}.
    The shaded regions represent the standard deviation from the same realisation corrected with 10 different parameters fitted from different initial conditions. }
    \label{fig:cross_appendix}
\end{figure}

As described in Sec.\ref{results}, we test the robustness of our correction scheme to changes in resolution and cosmological parameters. In the top panel of \autoref{fig:cross_appendix} we show the results for the cross correlation coefficients between the Camels and the PM field, both computed from simulation generated at the fiducial cosmology of $\Omega_m=0.10$. 
It can be seen that, while the cross correlation coefficients between Camels and PM improves after the NN correction, the improvement after the PGD correction is less evident. \\
The results for the cross correlation coefficients between the Camels and the PM field computed when increased the resolution of the simulation are shown in the bottom panel of \autoref{fig:cross_appendix}. 
We point out that the improvement after our scheme correction is better than PGD, although most of the correlation between the two different fields is not recovered on small scales.

\onecolumn

\end{document}